\documentclass{article}

\usepackage {graphicx,latexsym}
\usepackage{amsmath,amsthm}
\usepackage{amsfonts}
\usepackage{amssymb}
\usepackage[utf8]{inputenc}
\usepackage{authblk}
\usepackage{mathtools}

\usepackage{caption}
\usepackage{subcaption}

\newcommand{\R}{{\mathbb R}}  
\newcommand{\C}{{\mathbb C}}  
\newcommand{\Z}{{\mathbb Z}}

\newcommand{\PP}{{\mathrm P}}    
  
\newcommand{\cp}{{\gamma^c}}

\begin{document}

\title{
Macroscopic observables from the comparison of local reference systems
}
\author[1]{
Claudio Meneses
\footnote{e-mail: \ttfamily meneses@math.uni-kiel.de}
}
\author[2]{
José A. Zapata
\footnote{e-mail: \ttfamily zapata@matmor.unam.mx}
}
\affil[1]{
Mathematisches Seminar, Christian-Albrechts Universit\"at zu Kiel, 
Ludewig-Meyn-Str. 4, 24118 Kiel, Germany.
}
\affil[2]{
Centro de Ciencias Matemáticas, 
Universidad Nacional Autónoma de México, 
C.P. 58089, Morelia, Michoacán, México.
}

\date{}
\maketitle

\begin{abstract} 

Parallel transport as dictated by a gauge field determines a collection of local reference systems. 
Comparing local reference systems in overlapping regions leads to an ensemble of algebras of relational kinematical observables for gauge theories including general relativity. 
Using an auxiliary cellular decomposition, we 
propose a discretization of the gauge field based on a decimation of the mentioned ensemble of kinematical observables. The outcome is 
a discrete ensemble of local subalgebras of ``macroscopic observables'' characterizing a measuring scale. 
A set of 
evaluations of those macroscopic observables is called an 
extended lattice gauge field because it determines a 
$G$-bundle over $M$ (and over submanifolds of $M$ that inherit a cellular decomposition) 
together with a lattice gauge field over an embedded lattice. 
A physical observable in our algebra of macroscopic observables is constructed. 
An initial study of 
aspects of regularization and coarse graining, which are special to this description of gauge fields over a combinatorial base, is presented. 
The physical relevance of this extension of ordinary lattice gauge fields is discussed in the context of quantum gravity. 
\end{abstract}

%
%
%
%
%

\section{Introduction}
%

General relativity takes place in spacetimes modeled as differential manifolds without fixed set of coordinates or any other type of background reference system. 
Observers have to construct their local reference systems using the fields. It is appealing to use as fundamental fields for gravity a frame field and a gauge field. The frame field provides reference systems at tangent spaces, which in particular encodes a metric. This would be enough for special relativity, but general relativity needs more general local reference systems. 
We will see in Section \ref{LocRefSys} how 
the gauge field characterizing parallel transport along curves in spacetime naturally provides a collection of reference systems (usually called local trivializations) and 
transition functions among them. 
Additionally, the local invariant of the gravitational gauge field --its  curvature-- couples to the matter distribution. When focussing on reference systems, a ``nonzero gravitational field'' is one in which the reference systems determined by the gauge field and the induced transition functions have nontrivial dependence on the paths used to define them. 

We study gauge fields on a smooth manifold $M$; 
for example, $M$ may be spacetime. 
The physical interest in modeling spacetime-local situations, those that could take place in a laboratory during the time period in which an experiment takes place 
forces us to consider confined regions of $M$ with boundary and corners. 
We remark that the framework that we will provide is also applicable to situations in which $M$ is a Cauchy surface or any other manifold. 
The natural internal gauge group relevant for describing general relativity and for constructing models of quantum gravity 
is $SO(3,1)$ or its double cover; however, some frameworks 
use $SO(3)$ (or its double cover) as internal gauge group. 
On the other hand, matter fields of physical interest are described by gauge theories based on $SU(n)$. 
In what follows we will describe the framework for a general Lie group $G$ whenever it is possible and deal with particular cases only when the context demands it.

The comparison between local trivializations is encoded in gluing maps determined by the gauge field aided by some auxiliary structure (including a cellular decomposition $C$ of $M$). 
The evaluation of the gluing maps can be seen as an ensemble of kinematical observables (functionals of the gauge field) that 
determine the $G$-bundle over $M$ and its restriction to submanifolds of $M$ that inherit a cellular decomposition from $C$. 
In Section \ref{MacroObs}
we use an auxiliary cellular decomposition to decimate the mentioned ensemble kinematical observables of the gauge field. 
This decimation yields a discrete ensemble of local subalgebras. 
A measuring scale is defined by the mentioned discrete ensemble of subalgebras; consequently, we refer to the functionals in the algebra as 
``macroscopic observables.'' 
The evaluation of those macroscopic observables 
produces what we call an 
extended lattice gauge field because it determines a 
$G$-bundle over $M$ (and over submanifolds of $M$) 
together with a lattice gauge field over an embedded lattice. 

In early stages of this work, an important goal was 
to provide a framework amenable for treating gauge theories over discretized spacetimes compatible with 
the general boundary formulation of field theory championed by Oeckl \cite{GBF}, which provides a foundational backbone to the spin foam approach to gauge field theory. 
The mentioned ensemble of kinematical observables, introduced in Section \ref{MacroObs}, achieves this goal in a precise sense.

Lattice gauge theory is based on the kinematical observables arising from the parallel transport along the discrete set of paths fitting inside a lattice representing spacetime. Loop quantization can be interpreted as arising from a continuum limit of lattice gauge theories \cite{LQasContLim}. 
Even when this algebra of kinematical observables has lead to interesting physics, it cannot capture topological aspects of the $G$-bundle induced by the gauge field in the continuum. This follows from the fact that 
its spectrum 
--the space of lattice gauge fields-- is connected, while the set of 
connected components of the space of gauge fields in the continuum is in one-to-one correspondence with the set of isomorphism classes of $G$-bundles over $M$. 
This leads to the conclusion that some essential information has been left out in the discretization of the gauge fields. 

Consider the following analogy. 
In the study of real valued functions on a smooth manifold $M$ by means of a discretization based on decimating functions by evaluation on a prescribed discrete set of points, we would discover that topological aspects of $M$ are lost. The solution in this case is to replace the discrete set of points with a richer discretization of $M$ like a triangulation. This discretization is based on a discrete set of vertices {\em together with} relations defining links between pairs of vertices that are declared as neighbors and similar relations defining higher dimensional simplices. 
We can study real valued functions on $M$ replacing $M$ with a simplicial complex representing it while retaining all the topological properties of the manifold. 
In Section \ref{LocRefSys}, where we describe the $G$-bundle determined by the gauge field, we use certain multi-parametric families of paths on $M$ and see that they form simplicial complexes of paths in $M$. The 
zero-dimensional simplices in those simplicial complexes of paths correspond to the paths appearing in an embedded lattice which could be used to extract a lattice gauge field from the gauge field in the continuum. 

In Section \ref{MacroObs}, we introduce a discrete ensemble of local subalgebras extending the subalgebras whose evaluation yields ordinary lattice gauge fields. 
The spectrum of the ensemble of extended subalgebras 
--the space of possible evaluations-- consists of extended lattice gauge fields that characterize a $G$-bundle over $M$ and determine an ordinary lattice gauge field on a lattice embedded in $M$. 
The crucial ingredient of that extension is a richer discretization of the path groupoid of $M$. Apart from a discrete set of paths playing the role of vertices, we use the simplicial complexes of paths entering the description of the $G$-bundle determined by the gauge field in the continuum. 
%
%
A relational physical observable is exhibited in Section \ref{ARelObs}.

A purely combinatorial definition of extended lattice gauge fields is given in Section \ref{RegSect}.  
We give a definition of a groupoid of combinatorial paths and a groupoid homomorphism from paths in the continuum to combinatorial paths. 
We then define simplices of combinatorial paths corresponding to the simplices of paths in the continuum that are relevant in the construction of gluing maps, relating local trivializations of the bundle over cells related by inclusion. 
In this context we study regularization (in Subsection \ref{CommentsReg}) and coarse graining (in Section \ref{CoarseGrainingSect}).

The physical relevance of this extension of ordinary lattice gauge fields is discussed in the context of quantum gravity. 
To provide a starting point for the discussion, we 
briefly comment on the case of lattice gauge theory for chromodynamics. 
In quantum chromodynamics, the topological susceptibility is known to play a role in the calculation of the masses of hadrons. We however mentioned that lattice gauge fields were incapable of storing topological information leading to a puzzle. 
Lüscher gave an answer to this puzzle \cite{Luscher}. He proved that 
if we are only concerned with the continuum limit, and under the assumption that 
this scenario requires taking the bare coupling constant to zero, 
some topological properties of the gauge field are recovered in the limit. 
In Section \ref{Summary+Outlook} we give more details about this argument and its relevance in quantum gravity. 
Phillips and Stone provided an extension of a construction used by Lüscher in his work mentioned above that is valid for an arbitrary compact base space and compact Lie group \cite{PhillipsStone}.

A spin foam study of euclidian gravity and BF theory in two dimensions 
by Oriti, Rovelli and Speziale 
revealed that an extension of the lattice gauge field was essential for capturing the correct physics \cite{2dSF}. 
Another recent development in the context of three-dimensional gravity on bounded domains shows that fields with nontrivial winding numbers play a crucial role \cite{Winding3dB, Winding3dW}. 
Our definition of extended lattice gauge fields may be seen as a higher dimensional non-abelian generalization of the extension used in \cite{2dSF}. 
In Section \ref{Summary+Outlook} we argue that 
the same topological mechanism making the extension of the gauge field relevant for euclidian two-dimensional gravity makes it potentially relevant for euclidian and lorentzian four-dimensional gravity.

\section{Local reference systems and gluing from the gauge field}
\label{LocRefSys}
%

In general, there could be more than one spacetime region of interest, and the consistency between two descriptions of a phenomenon occurring in the intersection of two intersecting spacetime regions is crucial.
Moreover, since the reference systems are constructed using restrictions of the field to local subsystems, 
the compatibility of the two descriptions leads to observables measuring the field with respect to itself. 
This observation plays a crucial role in our work.

In this section we will show that a gauge field determines local reference systems, and 
in Section \ref{MacroObs} we will introduce an algebra of kinematical observables describing the gauge field. The definition of these two structures needs a cellular decomposition $C$ of $M$, as well as some further auxiliary structure described below. 
One reference for cellular decompositions and triangulations of manifolds is \cite{CellDecs}. 
Cells will be labeled by Greek letters. We will write 
$c_\nu \in C$, and this will mean a cell of the cellular decomposition. We can also use the same symbol for a subset of the manifold $c_\nu \subset M$. 
There are cells of all dimensions from $0$ to $n= {\rm dim}M$; 
the set of $k$-dimensional cells will be denoted by $C^k$. 
Two basic properties of a cellular decomposition are: (i) 
$M = \cup_C c_\nu$ and (ii) $c_\nu \cap c_\mu = \emptyset$ unless 
$\nu = \mu$. 

It is also true that $M$ is the union of the closure of cells of maximal dimension $M = \cup_{C^n} \bar{c}_\nu$. 
A cell of maximal dimension $c_\nu$ can be 
used to model an ``atom of spacetime,'' a minimal confined spatiotemporal region. 
We may consider the closed maximal dimension cells as retractions of elements of an open cover of $M$ in which the overlap has been minimized. 
We will work with a cellular decomposition $C$ that admits a refining triangulation ${\mathsf N}_C$. This type of cellular decompositions includes decompositions into hypercubes, triangulations and co-triangulations; in this type of cellular decompositions, the boundary of the closure of a $k$-cell is a $(k-1)$-sphere. 
A detailed description of the triangulation is the following:  
Each vertex of the abstract simplicial complex ${\cal N}_C$ 
corresponds to a subset of $M$ determined by a $k$-tuple intersection of the closures cells of $C^n$. 
One-dimensional simplices of ${\cal N}_C$ are formed by pairs of vertices $v_1, v_2$ with the property that the corresponding subsets of $M$ are related by inclusion, either $S(v_1) \supsetneq S(v_2)$ or 
$S(v_2) \supsetneq S(v_1)$. 
Similarly, a set of $k+1$ vertices determines a $k$-simplex of ${\cal N}_C$ if the corresponding subsets of $M$ can be ordered by strict inclusion. 
A consequence of this definition is that in a $k$-simplex of 
${\cal N}_C$ the set of vertices inherits an order. 
The auxiliary structure 
is completed by a choice of homeomorphism $\phi_C: |{\cal N}_C| \to M$ such that the image of each $k$-simplex is contained in the cell of $C$ that labels it. 
The pair composed by the abstract simplicial complex and the homeomorphism $({\cal N}_C , \phi)$ 
provides a triangulation 
${\mathsf N}_C$ of $M$ that refines $C$. 
In some situations it is appropriate to think of ${\mathsf N}_C$ as a barycentric subdivision of $C$. 
The $1$-skeleton of the refining triangulation 
gives us an embedded lattice ${{\mathsf L}_C} = {\mathsf N}_C^{(1)}$. 
Objects associated with higher dimensional simplices of ${\mathsf N}_C$ will be  relevant only up to certain relative homotopy relations.

\subsection{Reference systems from the gauge field}
In the spirit of Barrett and Kobayashi \cite{Barrett, Kobayashi},  
a $G$-gauge field is considered in terms of the parallel transport map which it induces. We will see that a gauge field characterizes a $G$-bundle over $M$ and a gauge orbit of connections in that bundle. See also the work of Lewandowski for a related study \cite{Lewandowski}. 
A more mathematically rigorous presentation of what is described in this section but lacks a physical interpretation can be found in \cite{ELGmath}. 

Intuitively, a gauge field gives a prescription for parallel transport along paths; however, since there are different paths such that every gauge field yields identical parallel transport, the notion of path needs to be refined. 

The path groupoid ${\cal P}_M$ consists of equivalence classes of piecewise smooth paths (or curves) in $M$ in which two paths are considered equivalent if they differ by a reparametrization or by retracing. 
The meaning of equivalence by retracing is that portions of piecewise smooth paths, 
which may be written as concatenations of the type 
--$\ldots \circ \gamma_3 \circ (\gamma_2)^{-1}\circ \gamma_2 \circ \gamma_1 \circ \ldots$--, are equivalent to portions of the type 
--$\ldots \circ \gamma_3 \circ \gamma_1 \circ \ldots$--; 
there are other related notions of path equivalence 
(see for example \cite{Barrett, CaetanoPicken}). 
After taking equivalence classes under reparametrization and retracing, the set of paths becomes a groupoid. 
For notational convenience we call the elements paths, and we will omit the brackets in $[\gamma]$.

The product, or composition, in the groupoid is defined as follows. 
Every path 
$\gamma \in {\cal P}_M$ has a source and a target $s(\gamma), t(\gamma) \in M$. Path $\gamma_1$ can be composed with path 
$\gamma_2$ if and only if $t(\gamma_1) = s(\gamma_2)$; the result of their composition is also denoted by $\gamma_2 \circ \gamma_1 \in {\cal P}_M$.

A reference system ${\cal F}_x$ is assigned 
to every point $x \in M$, 
and they can all be identified with a typical space ${\cal F}$. 
The standard terminology is to call ${\cal F}_x$ the fiber over $x$. 
The internal gauge group $G$ must be able to act on this collection of reference systems (with a global right action). 
It will simplify our framework to consider ${\cal F}\simeq G$; 
if a field of interest is naturally described using a typical space ${\cal F}$ which is a vector space, an affine space or a sphere, the work described in the rest of the article still provides valuable information regarding the bundle and the gauge field. 
A gauge field, usually denoted by $A$, can be considered the object providing a parallel transport map for every path. With the use of the conventions announced above, we could think of $A(\gamma)$ as proving a map 
\[
A(\gamma): {\cal F}_{s(\gamma)} \to {\cal F}_{t(\gamma)} ,  
\]
\[
A(\gamma) \triangleright g_s = g_t \in {\cal F}_{t(\gamma)}
\simeq G  . 
\]
Since the action must commute with the global right $G$ action on the fibers, our description would be simplified if each fiber were identified with $G$ because in that escenario $A(\gamma)$ could be represented by a group element acting by left multiplication. 
However, a continuous identification of the fibers with the internal gauge group may be possible only locally. 
One way to proceed is to only consider paths whose source and target belong to a discrete collection of base points and give, an arbitrary identification of the fiber over each of those base points with $G$. 
The discrete collection of base points that we will use is the set of vertices of the refining triangulation. For every cell $c_\nu \in C$ there is a corresponding vertex $p_\nu \in {\mathsf N}_C^0$, and we will fix an identification between ${\cal F}_{p_\nu}$ and $G$ for each base point in ${\mathsf N}_C^0$. We will write ${\cal F}_{p_\nu} \equiv G$. 
Accordingly, we will work with the path subgroupoid 
${\cal P}_{M, {\mathsf N}_C^0} \subset {\cal P}_M$, which consists of paths whose source and target belong to ${\mathsf N}_C^0$. 
When we use these conventions, we can write $A(\gamma) \in G$ and 
$A(\gamma) \triangleright g_s = A(\gamma) g_s$; 
the assignment of group elements to paths 
\[
{\cal P}_{M, {\mathsf N}_C^0} \overset{A}{\longrightarrow} G  
\]
is a groupoid homomorphism. 
This parallel transport homomorphism is invariant under any gauge transformation whose restrictions to ${\mathsf N}_C^0$ is the identity. 
It turns out that the parallel transport homomorphism 
completely characterizes gauge fields modulo the restricted gauge group fixing the set of fibers over ${\mathsf N}_C^0$. 
The gauge field would be called smooth if the parallel transport homomorphism meets a smoothness criterion to be described below. 


Now we will describe how 
the gauge field induces a $G$-principal fiber bundle over every closed cell $\bar{c}_\nu$. 

%

Let $p_\nu \in {\mathsf N}_C^0$ be the vertex of the refining triangulation corresponding to cell $c_\nu$. 
We will consider paths $[\gamma] \in {\cal P}_M$ starting at the base point $s(\gamma) = p_\nu$ and finishing somewhere in the cell 
$t(\gamma) = x \in \bar{c}_\nu$, which additionally have the property of having a representative completely contained in the cell $\gamma \subset \bar{c}_\nu$; the set of those paths will be denoted by 
$\widetilde{{\cal P}}_{c_\nu}$. 
The gauge field $A$ parallel transports initial conditions 
$g \in {\cal F}_{p_\nu} \equiv G$ 
at the base to any point $x = t(\gamma) \in \bar{c}_\nu$ 
along paths in $\widetilde{{\cal P}}_{c_\nu}$ to construct a fiber over $x$. The consistency conditions among descriptions of the fiber over $x$ obtained using different paths is written as an equivalence relation. 
In the set $G \times \widetilde{{\cal P}}_{c_\nu}$, two pairs 
$(g_1 , \gamma_1)$ and $(g_2 , \gamma_2)$ 
are defined to be equivalent if 
\begin{itemize}
\item
$t(\gamma_1) = t(\gamma_2)$ and  
\item
$A(\gamma_2^{-1} \circ \gamma_1) g_1 = g_2$. 
\end{itemize}
In the definition of equivalence classes, only 
paths of the form $\gamma_2^{-1} \circ \gamma_1$ participate; 
they are loops based on $p_\nu$, and they form a group 
${\cal L}_{c_\nu}$. 
This group is contained in the 
groupoid 
${\cal P}_{c_\nu} = \widetilde{{\cal P}}_{c_\nu} \cap 
{\cal P}_{M, {\mathsf N}_C^0}$, which will also be relevant in the following subsection. 
The gauge field $A$ restricted to $\bar{c}_\nu$ induces 
a groupoid homomorphism  
${\cal P}_{c_\nu} \overset{A}{\longrightarrow} G$. 
Notice that this groupoid homomorphism is trivial, leading to a set of equivalence classes that is independent of the path, only if the curvature of the gauge field vanishes.

The resulting set of equivalence classes 
$\PP_\nu = \PP_\nu (A) = (G \times \widetilde{{\cal P}}_{c_\nu})/\sim_A$ 
can be turned into a principal $G$-bundle over 
$\bar{c}_\nu$, with the projection map given by 
$\pi([g, \gamma]) = t(\gamma)$. The right $G$ action on the fibers 
$\pi^{-1}(x)$ is given by 
$h\triangleright [g, \gamma] = [gh, \gamma]$. It is clear that the right $G$ action induces a bijection (for each choice of auxiliary base point in the fiber) 
between $G$ and the typical fiber.

To see that $\PP_\nu$ is a principal $G$-bundle, we 
provide a trivialization determined by the gauge field with the aid of our auxiliary structure. 
In every closed cell $\bar{c}_\nu$, there is a path system joining 
its points to the base point: 
$x \mapsto \gamma_\nu^x \in \widetilde{\cal P}_{c_\nu}$ where the path starts at $s(\gamma_\nu^x) = p_\nu$, finishes at $t(\gamma_\nu^x) = x$, and it is a straight path according to the triangulation ${\mathsf N}_C$. 
The trivialization 
$G \times \bar{c}_\nu \overset{\varphi_\nu}{\longrightarrow} \PP_\nu$ 
is 
\[
(g, x)_\nu \overset{\varphi_\nu}{\longmapsto} [g, \gamma_\nu^x ] . 
\]
This map provides a smooth structure for $\PP_\nu$. 

Additionally, the described structure comes with a natural lift for curves in the base. Given any curve $c: [0,1] \to \bar{c}_\nu$ and an initial condition $[g, \gamma] \in \pi^{-1}(s(c))$, we will exhibit a curve in $\PP_\nu$ lifting the curve and starting at the initial condition. 
Let $c_t$ be the portion of the curve $c$ starting at $c(0) = t(\gamma)$ and finishing at $c(t)$. The lifting curve is 
$t \mapsto [g, c_t \circ \gamma]$.


Two smooth gauge fields $A, A'$ in $\bar{c}_\nu$ are considered to be gauge equivalent if there is a smooth $G$-valued function $g: \bar{c}_\nu \to G$ such that for any path 
$\gamma$ we have 
$A'(\gamma) = g(t(\gamma)) A(\gamma) g(s(\gamma))^{-1}$. 
It is interesting to see that in the construction of the bundle presented above a modification of the gauge field $A$ to the field $A'$ is only sensitive to the value of $g$ at 
$p_\nu$. 
There is a residual gauge dependence in the construction resulting in $A'$ inducing a bundle $\PP'_\nu$ with a curve lifting structure that is related to the one constructed above by the bundle map 
$[h, \gamma ] \mapsto [g(p_\nu)^{-1} h, \gamma ]'$. 
This happens because the construction of the bundle and the curve lifting map do not depend on the gauge field $A$, but on its orbit according to a subgroup of the gauge group whose restriction to $p_\nu$ 
is the identity. 
On the other hand, the construction of the local trivialization is much more arbitrary; it is sensitive to the choice of homeomorphism 
$\phi_C: |{\cal N}_C| \to M$ encoded in the triangulation.

We have seen that a gauge field $A$ thought of as a smooth parallel transport map gives us a smooth principal $G$-bundle, as well as a connection on it. 
The smoothness of the connection 
according to the smooth structure of $\PP_\nu$ 
provides a condition for the gauge field $A$ to be smooth. 
We will comment more on this subject in the next subsection. 

In order to finish the description of the gauge field, we have to glue together the collection of bundles $\{ \PP_\nu \}_{c_\nu \in C}$. Gluing is the subject of the next subsection.

The construction given above, which applies to the gauge field over a single cell $c_\nu$, is just a specialization of Barrett's construction \cite{Barrett} to the case where the manifold $M$ consists of a single cell. 
Our contributions start in the next subsection where 
we will describe the bundle and the curve lifting for the whole manifold $M$ by gluing bundles over the cells of the cellular decomposition. 
This cellularly local treatment will allow us to implement a decimation procedure in the following sections making contact with lattice gauge theories and spin foam models.

\subsection{Gluing local reference systems as dictated by the gauge field}
\label{GlLocRefSubSec}

We considered a spacetime $M$, endowed with a cellular decomposition $C$ and a refining triangulation ${\mathsf N}_C$ as auxiliary structures, 
and we saw that a $G$ gauge field $A$ leads to a collection of principal $G$-bundles $\PP_\nu \to \bar{c}_\nu$ over the closure of the cells in $C$ endowed with local trivializations. 
Recall that we may consider the closed maximal dimension cells as retractions of elements of an open cover of $M$ in which the overlap had been minimized, $M = \cup_{C^n} \bar{c}_\nu$. 
Thus, we can think of the local trivializations of the local bundles $\PP_\nu$ over maximal dimension cells as local reference systems. 

This collection of local reference systems has regions of multiple overlaps corresponding to the locus of lower dimensional cells. 
The translation between the multiple descriptions of the system in these regions will be provided by a set of gluing maps describing 
changes of local trivialization over pairs of nested closed cells 
$\bar{c}_\nu \supsetneq \bar{c}_\tau$.

Consider a pair of nested closed cells $\bar{c}_\nu \supsetneq \bar{c}_\tau$. The bundles 
$\PP_\nu|_{\bar{c}_\tau}$ and $\PP_\tau$ are naturally identified using the same principle leading to the equivalence relation defined in the previous subsection. Given any $x \in \bar{c}_\tau$ 
and a pair of paths in the respective closed cells satisfying 
$s(\gamma_\nu) = p_\nu$, $s(\gamma_\tau) = p_\tau$, 
$t(\gamma_\nu) = t(\gamma_\tau) = x$, 
we declare the following identification 
\[
[g, \gamma_\nu ]_\nu \sim_A 
[ A((\gamma_\tau)^{-1} \circ \gamma_\nu) g, \gamma_\tau ]_\tau . 
\]
Notice that the paths participating in the definition of this identification belong to the groupoid ${\cal P}_{c_\nu}$, but they are not loops; they start in $p_\nu$ and finish in $p_\tau$. 
The gauge field restricted to act on 
loops in ${\cal P}_{c_\nu}$ 
or ${\cal P}_{c_\tau} \subset {\cal P}_{c_\nu}$ 
is responsible for defining the bundles over $\bar{c}_\nu$ and $\bar{c}_\tau$, while the gauge field acting on open paths is responsible for the identification of the bundles in the overlapping regions. 

It is also interesting to describe the change of local trivializations of the bundle over $\bar{c}_\tau$ that may be induced using the auxiliary structure over $\bar{c}_\nu$ or over $\bar{c}_\tau$. Given any $x \in \bar{c}_\tau \subset \bar{c}_\nu$, we have 
\[
(g, x)_\nu \overset{\varphi_\nu}{\longmapsto} [g, \gamma_\nu^x ]_\nu  \sim_A 
[ A((\gamma_\tau^x)^{-1} \circ \gamma_\nu^x) g, \gamma_\tau^x ]_\tau \overset{\varphi_\tau^{-1}}{\longmapsto} 
( 
A((\gamma_\tau^x)^{-1} \circ \gamma_\nu^x)  g, x)_\tau . 
\]
In this way the gauge field determines a gluing map 
\[
g_{\tau \nu} [A] = A((\gamma_\tau^x)^{-1} \circ \gamma_\nu^x): 
\bar{c}_\tau \to G , 
\]
characterizing the change of local trivializations of nested subcells 
$\bar{c}_\nu \supsetneq \bar{c}_\tau$. 
Notice that the gluing map is constant when the curvature of the field vanishes. One may say that the nontriviality of the gluing map measures a certain integration of the curvature.

We can also consider a pair of neighboring $n$-dimensional cells sharing an $(n-1)$-dimensional cell 
$\bar{c}_\nu \cap \bar{c}_{\nu'} = \bar{c}_\tau$. The corresponding change of local trivialization is dictated by transition functions, which can be calculated from the gluing maps 
\[
\psi_{\nu \nu'}[A]  = 
g_{\tau \nu}^{-1}  \circ g_{\tau \nu'} [A] = 
A((\gamma_\nu^x)^{-1} \circ \gamma_{\nu'}^x) : 
\bar{c}_\tau \to G . 
\]

The result of gluing the bundles over all the cells in the cellular decomposition is a $G$-bundle $\PP_{M, A}$ over $M$, which may not globally be a product even when all its pieces over the cells are trivial. Over a given base there may be gauge fields leading to nonequivalent $G$-bundles. 
The bundle structure is encoded in the gluing maps given above.

\section{Effective gauge theories and macroscopic \\
observables}
\label{MacroObs}
%

In order to prescribe an effective theory at some ``macroscopic scale,'' we select a subalgebra of the algebra of observables of the system. That subalgebra models the measurements that are available to a given measuring setup, and it determines the measuring scale. 

Because for the moment we are ignoring the dynamics, 
when we write observables we mean a sort of kinematical observables where we only demand invariance under internal gauge transformations. 
In the next section we will exhibit one physical observable for gravity and chromodynamics in a very specific situation, and 
in Section \ref{Summary+Outlook} we will discuss 
the relevance of our framework when the dynamics of chromodynamics or quantum gravity are taken into account.

The choice of a subalgebra of observables is crucial. It corresponds to selecting an experimental setup, with the specific aspects of the field that the experimentalist wants to measure and with all the other aspects of the field that the experimentalist must control for the study to be well defined. 
At an intuitive level, a reasonable condition is that the evaluation of all the observables in the subalgebra determine the state of the field up to ``microscopic details.'' This is too vague because by definition those ``details'' are what the experimental setup cannot detect. Without detailed knowledge of the space of fields, we cannot make precise statements. 
We will base our study on considerations following from the description of the gauge field as providing parallel transport maps. We have seen that complete knowledge of the parallel transport maps determines the (restricted) gauge orbit of the field. 
Lattice gauge theory is based on considering parallel transport on paths contained in a discrete lattice. If those paths are pictured as embedded in a base where there is a gauge field in the continuum, the choice of a discrete lattice of allowed paths leads to partial knowledge of the gauge field. 
Could we say that we know the field up to ``microscopic details''? 

It is common to choose as macroscopic observables properties that are extensive in the sense that they are additive with respect to a spatial partition of the system into subsystems. 
Since gauge fields are intrinsically nonlinear, this property is not a natural criterion, but it would be fair to say that holonomies (parallel transport maps along closed paths) 
are extensive in a nonlinear way. A nonabelian Stokes theorem would tell us that the holonomy along a loop bounding a surface measures the 
``integration of the curvature on that surface.'' Thus, characters of holonomies (real valued functions depending on the conjugacy class) are gauge invariant ``extensive'' observables for the curvature. 

On the other hand, an important issue is that the evaluation of holonomies on a discrete collection of loops cannot tell us if the gauge field has big or small curvature. For example, in a case with $G = U(1)$, it is possible that the holonomies on all the loops contained in a discrete collection (for example, all the plaquettes in a given embedded lattice) evaluate to ${\rm id} \in U(1)$, but that if one considers a one-parameter family of loops interpolating between a point and one of the considered loops, the corresponding holonomies would wind around $U(1)$ several times. 
Another aspect of the same issue is that if $G$ is connected 
in lattice gauge theory the space of fields is path connected; 
any two lattice gauge fields can be connected by a curve. 
This is not the case in the continuum. In particular, we mentioned that in the continuum there may be gauge fields inducing inequivalent $G$-bundles over a given base $M$. This is possible only because the space of gauge fields in the continuum is not connected; otherwise there would be a continuous interpolation between inequivalent bundles over $M$. 

In the previous section, we described gluing maps 
$g_{\tau \nu} : \bar{c}_\tau \to G$ 
for the bundles induced by gauge fields in the continuum. 
The gluing maps encode the bundle structure, but 
at a given measuring scale we cannot store all the information contained in the gluing maps. 


After extracting gauge invariant information, we could consider the evaluation of gluing maps on vertices of ${\mathsf N}_C$ 
as macroscopic observables. The family of macroscopic observables 
\begin{equation}\label{glv}
\{ \{ g_{\tau \nu} [A] (p_\sigma) \in G\}_{\bar{c}_\sigma \subseteq \bar{c}_\tau} \} \mbox{ for all pairs of nested closed cells } \bar{c}_\nu \supsetneq \bar{c}_\tau 
\end{equation}
has three important properties: 
\begin{itemize}
\item
A {\em lattice gauge field} 
on the embedded lattice ${{\mathsf L}_C}= {\mathsf N}_C^{(1)}$ 
determines their evaluation. 
In addition, their evaluation 
generates a lattice gauge field on ${{\mathsf L}_C}$. The evaluations of 
(\ref{glv}) are not free generators; they are subject to relations generated by ``cocycle type'' relations 
$g_{\sigma \nu} (p_{\sigma'}) = g_{\sigma \tau} (p_{\sigma'}) g_{\tau \nu} (p_\sigma)$, 
which hold for 
any triple of nested closed cells 
$\bar{c}_\nu \supsetneq \bar{c}_\tau \supsetneq \bar{c}_\sigma$ and every 
$\bar{c}_{\sigma'} \subseteq \bar{c}_\sigma$. 

Viewing (\ref{glv}) as observables determining a lattice gauge field gives us an algebraically powerful tool. Consider the subgroupoid of 
the path groupoid whose elements have representatives fitting in the embedded lattice ${{\mathsf L}_C}$, and name it 
${\cal P}_{{\mathsf L}_C} \subset {\cal P}_M$. Thus, the evaluation of (\ref{glv}) characterizes a groupoid homomorphism from ${\cal P}_{{\mathsf L}_C}$ to $G$.

One consequence is that 
the family of gluing observables includes holonomies and their characters, which as mentioned above 
are the appropriate nonlinear versions of extensive observables measuring curvature. 
\item
Each observable has the interpretation of being a {\em relational observable}  
because it describes one local reference system constructed using the field with respect to another local reference system also constructed using the field. For the context of relational observables see \cite{PartObs} and 
\cite{RelObs} and references therein. 
\item
In contrast with the evaluation of the family of observables 
$\{ g_{\tau \nu} (x) \}$ for every $x \in \bar{c}_\tau$ and every 
$c_\tau \subset \partial c_\nu$, 
their evaluation does not characterize the gluing of the bundle over 
$\bar{c}_\nu$ to the bundle over its boundary, nor the total bundle over $M$. 
\end{itemize}

The last property means that extra macroscopic observables are needed to characterize the most basic qualitative feature of the gauge field. 
Since the gluing maps $g_{\tau \nu}$ determine the bundle structure, we should be able to find 
the missing macroscopic observables 
as the essential missing information from the evaluation of $g_{\tau \nu}$ only on 
vertices of ${\mathsf N}_C|_{\bar{c}_\tau}$. 

In \cite{ELGmath} we use a smaller embedded lattice 
that does not record the evaluation of the gluing map $g_{\tau \nu}$ at 
$p_\tau$, but only records the evaluation at vertices of 
$\bar{c}_\tau$. Here we use this version of ${{\mathsf L}_C}$ because it makes the study presented in Sections 
\ref{RegSect} and \ref{CoarseGrainingSect} simpler. 
In the Appendix, we give a map assigning an extended lattice gauge field as defined in \cite{ELGmath} to every extended lattice gauge field as defined in the article.

Below we will define new macroscopic observables to complement the family  presented in (\ref{glv}).

%
%
%

Given a pair of nested closed cells of $C$, 
$\bar{c}_\nu \supsetneq \bar{c}_\tau$, 
we have a gluing map $g_{\tau \nu}: \bar{c}_\tau \to G$; 
if ${\rm dim}(c_\tau) \geq 1$, 
its restriction to simplices of ${\mathsf N}_C$ 
contained in $\bar{c}_\tau$ are of interest to us. 
We denote 
$k$-dimensional simplices of the refining triangulation by 
${\mathsf s}^k \in {\mathsf N}_C^k$; 
the restriction of the gluing map 
to a $k$-simplex is written as 
$g_{\tau \nu}|_{{{\mathsf s}^k}}: {{\mathsf s}^k} \to G$. 

Now we will complement the set of macroscopic observables (\ref{glv}). 
To each pair of nested closed cells $\bar{c}_\nu \supsetneq \bar{c}_\tau$, 
of dimensions $k $ and $l\geq 1$ respectively, 
we associate a family of 
{\em gluing extension type} observables defined as follows: 
\begin{enumerate}
\item 
To each 
${\mathsf s}^1 \in {\mathsf N}_C^1$ such that  
${{\mathsf s}^1} \subset \bar{c}_\tau$
we associate the observable 
\begin{equation}\label{gl}
[ g_{\tau \nu} [A] |_{{{\mathsf s}^1}} ] \in {\mathrm GlExt}_1 , 
\end{equation}
which is the functional of $A$ that yields the homotopy class of the 
curve $g_{\tau \nu} [A]({\mathsf s}^1) \subset G$ 
relative to fixed points at 
$g_{\tau \nu} [A](\partial {{\mathsf s}^1})$ 
determined by the evaluation of (\ref{glv}). 
\item
If $l \geq 2$ 
to each ${\mathsf s}^2 \in {\mathsf N}_C^2$ such that 
${{\mathsf s}^2} \subset \bar{c}_\tau$ 
we associate the observable\\
\begin{equation}\label{gl2}
[ g_{\tau \nu} [A]|_{{{\mathsf s}^2}} ] \in {\mathrm GlExt}_2 , 
\end{equation}
which is the functional of $A$ that yields the homotopy class of the 
surface with boundary and corners 
$g_{\tau \nu} [A]({\mathsf s}^2) \subset G$ 
relative to fixed points at 
$g_{\tau \nu} [A](\partial {{\mathsf s}^2})$ 
determined by the evaluation of (\ref{glv}), 
and also constrained to induce the value of 
$[ g_{\tau \nu} [A]|_{{{\mathsf s}^1}} ] \in 
{\mathrm GlExt}_1$ 
for all 
${\mathsf s}^1 \subset \partial {{\mathsf s}^2}$ 
determined by the evaluation of (\ref{gl}). 

The process of increasing the dimension of the simplices 
${\mathsf s}^m$ continues until $m = l = \dim(c_\tau)$. 

\item[$m$.]
The gluing extension observables associated to each dimension $m$ simplex 
${\mathsf s}^m \in {\mathsf N}_C^m$ such that 
${{\mathsf s}^m} \subset \bar{c}_\tau$ 
is 
\begin{equation}\label{gll}
[ g_{\tau \nu} [A]|_{{{\mathsf s}^m}} ] \in {\mathrm GlExt}_m , 
\end{equation}
which is the functional of $A$ that yields the homotopy class of the 
$m$-dimensional hypersurface with boundary and corners 
$g_{\tau \nu} [A]({\mathsf s}^m) \subset G$ 
relative to fixed points at 
$g_{\tau \nu} [A](\partial {{\mathsf s}^m})$ 
determined by the evaluation of (\ref{glv}), 
and also constrained to induce the value of 
$[ g_{\tau \nu} [A]|_{{{\mathsf s}^k}} ] \in 
{\mathrm GlExt}_k$ 
for all 
${\mathsf s}^k \subset \partial {{\mathsf s}^m}$ 
determined by the evaluation of (\ref{gll}) for all values of 
$k \in \{ 1, \ldots , m-1 \}$. 
\end{enumerate}
Below we give a list of remarks about the family of observables introduced above to complement the family of observables (\ref{glv}). In the first remark we provide elements for better understanding of 
the sets ${\mathrm GlExt}_m$ 
in the case of a few groups of potential interest. 
\begin{itemize}
\item
It is a discrete family of observables valued in a discrete set. 
Each of the sets ${\mathrm GlExt}_m$ is discrete. Before discussing the set of possible evaluations of the family of observables 
(\ref{gll}) 
on a given gauge field, we give a brief description of 
the set ${\mathrm GlExt}_m$ of relative homotopy classes of 
$m$-dimensional 
hypersurfaces with boundary and corners in $G$ 
in particular examples. 
It is a discrete set on which $\pi_l (G)$ acts transitively and freely. 
Here is a list of homotopy groups of a few groups of potential interest: 
$\pi_1(U(1)) = \Z$, $\pi_1(SO(3)) = \pi_1(SO(4))= \pi_1(SO^+(3,1)) = \Z_2$, $\pi_1(SU(2)) = \pi_1(SU(3)) = \pi_1(SL(2, \C)) = 0$. $\pi_2 (G) = 0$ for 
all Lie groups. 
$\pi_3(U(1)) = 0$, $\pi_3(SO(3)) = \pi_3(SO^+(3,1)) = \pi_3(SU(2)) = 
\pi_3(SU(3)) = \pi_3(SL(2, \C)) = \Z$, and 
$\pi_3(SO(4)) = (\Z)^{\times 2}$. 

The set of possible evaluations of the observables defined above is not free; it is constrained to satisfy a set of compatibility conditions. 
Given a fixed set of evaluations of (\ref{glv}), 
there is a discrete group $R_C(G)$ acting on the set of possible evaluations 
$\{ [ g_{\tau \nu} |_{{{\mathsf s}^k}} ] \in {\mathrm GlExt}_k \}$ 
with the property that given any two 
gauge fields $A, A'$ for which the evaluation of (\ref{glv}) coincides, 
the sets of evaluations 
$\{ [ g_{\tau \nu}|_{{{\mathsf s}^k}} ] [A]\}$ and 
$\{ [ g_{\tau \nu}|_{{{\mathsf s}^k}} ] [A']\}$ 
are related by a unique 
$r \in R_C(G)$ leading to 
$r \triangleright \{ [ g_{\tau \nu}|_{{{\mathsf s}^k}} ] [A]\} = 
\{ [ g_{\tau \nu}|_{{{\mathsf s}^k}} ] [A']\}$. 
The group is constructed from a set of local generators subject to local relations. There is a generator for each simplex of ${\mathsf N}_C$ 
contained in the smallest cell of a 
pair of nested closed cells of $C$
$\bar{c}_\nu \supsetneq \bar{c}_\tau \supset {{\mathsf s}^k}$; 
a copy of the group $\pi_k (G)$ acts on the evaluation 
$[ g_{\tau \nu}|_{{{\mathsf s}^k}} ]$ changing it to any other value. 
The relations among the generators imply that for given any element 
$r \in R_C(G)$ (and for any gauge field $A$) the set of values 
$r \triangleright \{ [ g_{\tau \nu}|_{{{\mathsf s}^k}} ] (A)\}$ is a plausible evaluation for a gauge field. This follows from two types of conditions. 
Conditions of the first type demand that 
given any simplex ${\mathsf s}^k$ with $k \leq n-1$ 
the induced 
homotopy type for $[ g_{\tau \nu}|_{\partial {{\mathsf s}^k}} ]$, 
calculated by gluing%
\footnote{
Some explanation about the gluing operation will be given in Section 
\ref{CoarseGrainingSect} and a detailed description is given in 
\cite{ELGmath}. 
}
 the data 
$\{ g_{\tau \nu}|_{{{\mathsf s}^m}} \}_{{\mathsf s}^m \subset \partial {{\mathsf s}^k}}$, 
says that the gluing map can be extended from 
$\partial {{\mathsf s}^k}$ to ${{\mathsf s}^k}$. 
Conditions of the second type are responsible for the 
extendibility to the interior of simplex ${{\mathsf s}^k}$ of the 
equation 
$g_{\sigma \nu}|_{{{\mathsf s}^k}} (x) = 
g_{\sigma \tau}|_{{{\mathsf s}^k}} (x) g_{\tau \nu}|_{{{\mathsf s}^k}} (x)$ (which according to the recorded data by evaluation of (\ref{glv}) 
holds for all the vertices of ${{\mathsf s}^k}$ for all subsimplices 
of the smallest cell in the triple of nested closed cells $\bar{c}_\nu \supsetneq \bar{c}_\tau \supsetneq \bar{c}_\sigma$). 
In other words, when starting with data 
$\{ [ g_{\tau \nu}|_{{{\mathsf s}^k}} ] 
\in {\mathrm GlExt}_k \}$ 
such that the above equation is extendible to 
the interior of ${{\mathsf s}^k}$, as is the case when the data comes from a gauge field, 
the condition is that 
$r \triangleright \{ [ g_{\tau \nu}|_{{{\mathsf s}^k}} ] \}$ 
continues to enjoy of the extendibility property. 
\item
The observables are ``extensive.'' 
We will give more details about this property in Section \ref{CoarseGrainingSect}, where we describe the coarse graining of the gauge field. 
The set of observables 
$\{ [ g_{\tau \nu}|_{{{\mathsf s}^k}} ] \}$ corresponding to all the subsimplices ${{\mathsf s}^k}$ of a given cell $\bar{c}_\tau$ can be glued to 
determine the homotopy type 
$[ g_{\tau \nu}]$. 
Moreover if the cellular decomposition $C$ has a refinement $C'$, 
the evaluation of the set of observables 
(\ref{gll}) corresponding to the $C'$ 
determines uniquely 
the outcome of the evaluation of the set of observables corresponding to $C$; in Section \ref{CoarseGrainingSect} we will describe the subject in greater detail, and in Section \ref{ARelObs} we will describe one example. At the moment we have studied several examples successfully, but we feel that understanding the general case 
is within reach. We expect to report on advances on this front in the near future. 
Additionally, 
in the examples that we have studied, we have seen that the group $R_C(G)$ can be calculated from $R_{C'}(G)$ by the addition of the corresponding local generators while preserving the relations. 
\item
Together with the usual lattice gauge theory data, which can be interpreted as the evaluation of gluing observables (\ref{glv}), this data set characterizes the bundle structure at a local level. 
The bundle over each $k$-cell is trivial, and the changes of local trivializations in overlapping regions 
$\psi_{\nu, \nu'} = g_{\tau \nu}^{-1}  \circ g_{\tau \nu'}$ are known up to relative homotopy. 
This characterizes the $G$-bundle over $M$ and also its restriction to any submanifold that inherits a cellular decomposition from $C$. 
\item
The setting just described provides an assignment of a local subalgebra of macroscopic observables 
${\rm Obs}_{R, C}$ 
to any $R$ region (or submanifold, possibly with boundary and corners) 
of $M$ which inherits a cellular decomposition from $C$. 
The assignment is  
such that (i) $R' \supseteq R$ implies 
${\rm Obs}_{R', C} \supseteq {\rm Obs}_{R, C}$, and 
(ii) ${\rm Obs}_{R, C}$ is generated by the collection of subalgebras 
${\rm Obs}_{\bar{c}_\tau, C}$ for the collection of cells of dimension 
${\rm dim} (R)$ and contained in $R$. 
\end{itemize}

The algebra of macroscopic observables ${\rm Obs}_{M, C}$ 
generated by the families (\ref{glv})-(\ref{gll}) 
is determined by a lattice ${{\mathsf L}_C}$ embedded in $M$, 
the cellular decomposition $C$ and its refining triangulation 
${\mathsf N}_C$. 
Since the family of observables described above have homotopical character, the detailed structure contained in $C$ and ${\mathsf N}_C$ is only relevant up to homotopy relative to ${{\mathsf L}_C}$.

The point of view stated in the introductory section is that the algebra 
${\rm Obs}_{M, C}$ determines a measuring scale. 

The set of possible evaluations of ${\rm Obs}_{M, C}$ extracts an 
{\em extended lattice gauge field} $A^c$ from a gauge field in the continuum: 
\[
A \xmapsto{\tiny \mbox{e-Obs}_C}{} A^c .
\] 
In Section \ref{RegSect} we will introduce 
extended lattice gauge fields without appealing to the preexistence of a gauge field in the continuum.

\section{A relational physical observable}
\label{ARelObs}

In the previous section, we introduced an ensemble of local subalgebras generated by certain kinematical observables, ${\rm Obs}_{R, C}$, based on a heavy auxiliary structure dressing the region of interest $R$ of 
the base manifold $M$. 
The functionals in that algebra are not in general physical observables because their dependence on the auxiliary structure leads to incompatibilities with the symmetries of the theory. 
In this short section, we work on a four-dimensional manifold $M$ with a $G$-gauge field such that $\pi_3(G)$ is not zero and 
extract a nontrivial observable from ${\rm Obs}_{R, C}$
that is invariant under internal gauge transformations and under diffeomorphisms of the base manifold.%
\footnote
{
A diffeomorphism $\phi: M \to M$ acts on the gauge field shifting the path 
$(\phi \triangleright A) (\gamma) = A (\phi^{-1} \gamma)$. 
} 
Below we will assume that $M \simeq S^4$, but since the observable measures the field only in a confined region of interest, the observable, defined in the sense of \cite{GaugeFromH+Obs}, applies to confined regions of any manifold.

Our assumption regarding the field and the auxiliary structures outside $R$ is that they provide a reference system 
(trivialization of the bundle) over $\partial R\simeq S^3$. 
In the case when $M \simeq S^4$, the construction described in Section \ref{LocRefSys} using a single $4$-cell in $R^c$ endowed 
with a system 
$\{ \gamma_{R^c}^x \}_{x \in  \partial R^c}$ of paths with 
$s(\gamma_{R^c}^x) = p_{R^c}$ and $t(\gamma_{R^c}^x) = x$ 
yields the desired reference system.

Consider a situation in which the system of interest is 
the field in a confined domain $R\subset M$, and that system will be measured with respect to a reference system provided by the field outside of $R$. 
The field outside of $R$ is assumed to be known and to provide a reference. 
Let the domain of interest be modeled by 
one of the cells of maximal dimension, $R= \bar{c}_\nu$. 
We saw how the gauge field, together with the auxiliary structure, induces a local trivialization on $\bar{c}_\nu$. 
Once the trivialization of the bundle over $\partial R$ has been constructed using the gauge field in the region of interest, 
the change of local trivialization 
$\psi_{c_\nu^c \bar{c}_\nu} [A]$ (where $c_\nu^c$ denotes the complement of $c_\nu$) is a functional of 
the field in the region of interest; we will write 
$\psi_{c_\nu^c \bar{c}_\nu} [A|_{\bar{c}_\nu}]$. 
Below we will see that evaluation of the observables 
in ${\rm Obs}_{\bar{c}_\nu, C}$ determines 
the change of local trivialization 
$\psi_{c_\nu^c \bar{c}_\nu} [A|_{\bar{c}_\nu}]: 
\partial c_\nu \simeq S^3 \to G$ 
up to relative homotopy. In turn, 
considering any representative of this 
relative homotopy class, and calculating its 
(unrestricted) homotopy class, leads to 
$[\psi_{c_\nu^c \bar{c}_\nu} [A|_{\bar{c}_\nu}] ] 
\in \pi_3(G) $; we will write 
\[
[\psi_{c_\nu^c \bar{c}_\nu}] [A^c|_{\bar{c}_\nu}] \in \pi_3(G) . 
\]
%
Let us describe how the extended lattice gauge field determines 
the relative homotopy class of the smooth map 
$\psi_{c_\nu^c \bar{c}_\nu}[A|_{\bar{c}_\nu}]$.
The restriction of $\psi_{c_\nu^c \bar{c}_\nu}$
to any of the closed $3$-cells in the boundary 
$\bar{c}_\tau \subset \partial c_\nu$, 
may be written as a product of gluing maps 
$\psi_{c_\nu^c \bar{c}_\nu}|_{\bar{c}_\tau} [A|_{\bar{c}_\nu}] = 
g_{\tau c_\nu^c}^{-1} \circ (g_{\tau \nu} [A|_{\bar{c}_\nu}])$. 
Now we recall that, by assumption, the field outside of $R$, together with any required auxiliary structures, determine the gluing map $g_{\tau c_\nu^c}$. 
The gauge field inside $R$, 
on the other hand, determines the collection of maps 
$\{ g_{\tau \nu} [A|_{\bar{c}_\nu}] 
\}_{\bar{c}_\tau \subset \partial c_\nu}$, 
and the relative homotopy classes of maps in this collection 
(restricted to simplices of the refining triangulation) 
are determined by the extended lattice gauge field. 
Thus, the extended lattice gauge field determines the relative homotopy class of maps in the collection 
$\{ \psi_{c_\nu^c \bar{c}_\nu}|_{{{\mathsf s}^3}}[A|_{\bar{c}_\nu}] 
\}_{{{\mathsf s}^3} \subset \bar{c}_\tau \subset \partial c_\nu}$. 
Moreover, the classes in the collection have matching boundary conditions, implying that 
they can be glued together to yield the relative homotopy class of the map 
$\psi_{c_\nu^c \bar{c}_\nu} [A|_{\bar{c}_\nu}]$. 
Since the relative homotopy classe 
is determined by the extended lattice gauge field $A^c|_{\bar{c}_\nu}$ 
we have arrived to the desired result.

From its construction it is clear that the functional is an example of a relational observable in the sense of Rovelli \cite{PartObs, RelObs}. 
The functional is 
determined by the evaluation of ${\rm Obs}_{\bar{c}_\nu, C}$, and it is clearly invariant under internal gauge transformations. Below we give an argument showing that it is also invariant under the relevant subgroup of diffeomorphisms of the base. 

In the physical situation that we are considering where there is a region of interest $R$ in which the field is being measured with respect to the field in $R^c$, the diffeomorphisms that should be regarded as gauge are those fixing the reference. Thus, we need to consider 
the subgroup ${\rm Diff}_{R \, \partial R}$ 
of diffeomorphisms from $R$ to itself which restrict to the identity in 
$\partial R$. For a more ample discussion on the subject, see \cite{GaugeFromH+Obs}. 
Generic transformations in ${\rm Diff}_{R \, \partial R}$ do not preserve the auxiliary structure used to construct the subalgebra 
${\rm Obs}_{\bar{c}_\nu, C}$. 
The claim that the functional 
$[\psi_{c_\nu^c \bar{c}_\nu} ] [A^c|_{\bar{c}_\nu}]$ is an invariant of this action follows immediately from the fact that the action is continuous, while $[\psi_{c_\nu^c \bar{c}_\nu} ] [A^c|_{\bar{c}_\nu}]$ is valued in the discrete set $\pi_3(G)$.

In the context of extended lattice gauge fields, our statement that 
$[\psi_{c_\nu^c \bar{c}_\nu} ] [A^c|_{\bar{c}_\nu}]$ is a physical observable, in the sense described above, is a basic statement. We would like to stress the fact that this observable cannot be described using ordinary lattice gauge fields. 

Different approaches to quantum gravity use gauge fields with different internal groups. The third homotopy group of some of them is: 
$\pi_3(SO^+(3,1)) = \pi_3(SO(3)) = \pi_3(SU(2)) = \pi_3(SL(2,\C)) = \Z$.

The topological charge, defined for a gauge field over any $4$-manifold $M$ in terms of the integral 
$Q= \frac{1}{4 \pi^2} \int_M {\rm Tr} (F \wedge F)$ coincides with the 
observable $[\psi_{c_\nu^c \bar{c}_\nu} ] [A|_{\bar{c}_\nu}]$ defined above when $M \simeq S^4$. 
Within the context of extended lattice gauge fields, the integral $Q$ can be regularized; in Section \ref{RegSect} we discuss the issue, and we will present a detailed study about it in the future. 
We mentioned that the observable 
$[\psi_{c_\nu^c \bar{c}_\nu} ] [A^c|_{\bar{c}_\nu}]$ 
may also be defined in base manifolds different from $S^4$; in those cases the value of the observable is not the topological charge.

If the region of interest $R$ has the topology of a closed disc, but it consists of more than one cell of maximal dimension, the 
calculation of $[\psi_{R^c R} ] [A^c|_{R}]$ would involve coarse graining of the extended lattice gauge field. 
In Section \ref{CoarseGrainingSect} we discuss the evaluation of the topological charge by means of coarse graining.

\section{Extended lattice gauge fields as gauge fields on a combinatorial base}
\label{RegSect}
%
There is a gauge theory defined over a base $M$ in the continuum, and here we will define a gauge theory defined over a base with purely  combinatorial structure ${\cal N}_C$.
Regularization takes kinematical observables from the continuum and assigns them observables defined on the gauge theory over a combinatorial base.

We saw that in the continuum the evaluation of parallel transport observables completely characterizes 
a $G$-bundle over $M$ and a connection on that bundle (modulo gauge equivalence). 
We will 
discuss the regularization for a family of parallel transport observables corresponding to a family of paths that is large enough to characterize the gauge field but does not include a class of paths which would make our framework technically cumbersome. 

The first step in our construction will be an assignment of combinatorial paths to paths in the continuum: 
\[
\gamma \mapsto \gamma^c . 
\]
In the case of loops, we will write $l \mapsto l^c$. 
In addition, we will also talk about deformations of paths. We will define 
combinatorial counterparts of multi-parameter families of paths. 
An extended lattice gauge field on the combinatorial base 
will assign group elements to combinatorial paths and other objects to $k$-dimensional simplices of combinatorial paths 
corresponding to certain $k$-simplices of paths in the continuum.

\subsection{Combinatorial paths from paths in the continuum}
Since we are considering at the starting point piecewise smooth paths in $M$, among the family of allowed paths there are paths containing smooth segments $r$ intersecting a codimension one cell $c_\tau \in C$ wildly. This means that the set of intersection points $r \cap c_\tau$ is an infinite sequence. 
We will not regularize parallel transport along such paths. 

Every point $x \in M$ is contained in a single $C$ cell, $x \in c_\nu \in C$. 
Paths intersecting the cells of $C$ in a tamed way 
can be assigned a locally finite sequence of cells in $C$. 
The sequences that can be obtained have the property that 
the closure of 
consecutive cells are related by inclusion: for the sequence 
$\{ \ldots , \nu , \nu' , \ldots \}$ 
either 
$\bar{c}_{\nu} \supset \bar{c}_{\nu'}$ or $\bar{c}_{\nu} \subset \bar{c}_{\nu'}$. 
In the abstract simplicial complex ${\cal N}_C$ described in the beginning of Section \ref{LocRefSys}, vertices are 
cells of $C$, and links are 
pairs of cells whose closure is related by inclusion. 
Thus, sequences of the type given above are simplicial paths in 
${\cal N}_C$. 
We will refer to them as combinatorial paths and denote them by $\tilde{\cp}$. 

Combinatorial paths have an orientation but do not have a parametrization. There is a very natural retracing relation among combinatorial paths. A retracing move acting on a path segment changes a segment of the form 
$\{ \ldots , \nu , {\nu'} , \nu , \ldots \}$ to the segment 
$\{ \ldots , \nu , \ldots \}$. 
Two combinatorial paths $\tilde{\cp}, \tilde{\cp}'$ are retracing equivalent if each of them can be connected to a third path $\tilde{\cp}''$ by a finite sequence of retracing moves. 
In the rest of the article, we will call retracing equivalence classes of combinatorial paths simply combinatorial paths, and if there is no danger of confusion we may refer to them as paths;  
they will be denoted by $\cp \in {\cal P}_C$. 
Notice that ${\cal P}_C$ is a groupoid that has a natural identification (an 
isomorphism) with the groupoid of paths in the continuum that fit in the embedded lattice  ${{\cal P}_{{\mathsf L}_C}}$ defined in the previous section.

Consider a path $\tilde{\gamma}$ in the continuum intersecting $C$ in a tamed way. Let us call $\gamma \in {\cal P}_M$ its equivalence class modulo reparametrization and retracing; we may write 
$\tilde{\gamma} \in \gamma$. 
There may be paths in $\gamma$ which intersect $C$ in a wild manner. We will delete elements of the classes of paths with this type of behavior; we will not even consider giving them a second chance. 
Consider two continuum paths in the same class and intersecting $C$ in a tamed way, $\tilde{\gamma}, \tilde{\gamma}' \in \gamma$. 
Since the induced 
combinatorial paths belong to the same class 
$\tilde{\cp}, \tilde{\cp}' \in \cp$, we can say that a continuum path 
$\gamma \in {\cal P}_M$ of this type induces a combinatorial path: 
$\cp = {\mathsf T}_C (\gamma) \in {\cal P}_C$. 
A good property of the assignment 
${\mathsf T}_C:{\cal P}_M \to {{\cal P}_C}$ just defined 
is that it is a groupoid homomorphism. We will also write 
${\mathsf T}_C:{\cal P}_M \to {{\cal P}_{{\mathsf L}_C}}$.

This is the first step in providing a combinatorial basis for a gauge theory over a discretized base. It corresponds to starting with a manifold $M$ and dividing it into regions using an open cover, for example. We could then try to regularize functions on $M$ as a set of values corresponding to averages on the regions (or 
corresponding to the set of evaluations on 
a collection of points representing each region). 
We know, however, that this procedure would destroy the topological properties of $M$ and that we would end up with a very poor combinatorial theory. A much better discretization procedure can be achieved by using a triangulation or cellular decomposition of $M$. 

This assignment of combinatorial paths to continuum paths was introduced in a different context in \cite{Cflat}. The assignment can be described in terms of how a path in the continuum intersects the set of closed codimension one cells. Viewed in this way, there is a clear resemblance to the coarse graining mechanism proposed by Livine in the context of canonical loop quantum gravity \cite{EteraCoarse}. See Section \ref{CoarseGrainingSect} for comments on this relation.

\subsection{Simplicial families of combinatorial paths 
from\\ 
multiparametric families of paths in the continuum}
Below we will give extra structure to enrich our discretization of the path groupoid. 
Consider a one-parameter family of paths in the continuum $\gamma_t \in {\cal P}_M$. 
The corresponding family of combinatorial paths 
$\cp_t = {\mathsf T}_C (\gamma_t) \in {\cal P}_C$ will be a discrete family in which there could be transitions from one combinatorial path to another one, which we know should be considered as a neighboring path. 
%
In the construction of the local trivializations and gluing maps given in the previous sections, $k$-parameter families of paths (with $k \in \{ 1, \ldots n-1\}$) played a crucial role; in this subsection we will provide a combinatorial counterpart of those families. 

The first step is to study the way in which $k$-parameter families of paths appear in formulas (\ref{gl}), (\ref{gl2}), (\ref{gll}) 
for the relative homotopy types of 
the gluing maps $g_{\tau \nu}$ for a fixed pair of nested closed cells 
$\bar{c}_\nu \supsetneq \bar{c}_\tau$ in $C$. 
The relevant families of paths in the continuum, 
$\{ \gamma_{\tau \nu}^x\}_{x \in {{\mathsf s}}}$, 
are labeled by simplices in ${\mathsf N}_C$ contained in $\bar{c}_\tau$; 
for any point $x \in {{\mathsf s}}$ there is a path 
$\gamma_{\tau \nu}^x = (\gamma_\nu^x)^{-1} \circ \gamma_\nu^x$ 
with source $p_\nu$ and target $p_\tau$. 
If two of these simplices are related by inclusion, 
${\mathsf s'}\subsetneq {\mathsf s}$ the corresponding path families are also related by inclusion. 
The sets of path families appearing in the calculation of 
gluing extension observables (\ref{gl}), (\ref{gl2}), (\ref{gll}) 
form a simplicial complex. There is a simplex of paths in the continuum for each simplex 
in the triangulation of $\bar{c}_\tau$ induced by 
${\mathsf N}_C$ 
(for each cell $c_\nu$ such that $\bar{c}_\nu \supsetneq \bar{c}_\tau$); in some sense what we have is a cell of paths in the continuum corresponding to 
the pair of nested closed cells $\bar{c}_\nu \supsetneq \bar{c}_\tau$. In \cite{ELGmath} these cells are not subdivided into subsimplices.

To each $k$-simplex of paths in the continuum 
determined by $\bar{c}_\nu \supsetneq \bar{c}_\tau \supset {\mathsf s}$, we will assign a $k$-simplex of combinatorial paths labeled by the corresponding objects in the abstract simplicial complex ${\cal N}_C$. 
This assignment follows from the 
natural isomorphism between the simplicial complexes 
${\cal N}_C$ and ${\mathsf N}_C$. 
The association of an abstract simplex to a simplex of the triangulation will be written as $s({\mathsf s})$. 
We will write $V(s)$ for the set of vertices of a simplex, which in this case consists of cells of $C$.
%
%
%
The assignment of the simplex of combinatorial paths to the simplex of paths in the continuum is 
\begin{equation}\label{spaths}
\{ \gamma_{\tau \nu}^x \}_{{x \in{\mathsf s}}} \mapsto 
\Gamma_{\tau \nu}^c(s) \doteq 
\{ \gamma_{\tau \nu}^{c \, \sigma} \}_{\sigma \in V(s({\mathsf s}))} , 
\end{equation}
where 
the definition of the combinatorial paths in the sets is simply 
$\gamma_{\tau \nu}^{c \, \sigma} = {\mathsf T}_C 
(\gamma_{\tau \nu}^{x= p_\sigma})$. 
Given the isomorphism between ${\mathsf N}_C$ and ${\cal N}_C$, it is clear that for a fixed pair of nested closed cells 
$\bar{c}_\nu \supsetneq \bar{c}_\tau$ 
the resulting families of sets of paths have the structure of the 
simplicial complex. 

The family of abstract simplices of combinatorial paths 
defined in (\ref{spaths}) 
is the extra structure that we will use to define a gauge field over a combinatorial base.

\subsection{Extended lattice gauge fields}
\label{ELGsubsec}

An {\em extended lattice gauge field} 
assigns different types of objects to abstract simplices of combinatorial paths $\Gamma_{\tau \nu}^c(s)$ of different dimensions. 
For $0$-simplices of combinatorial paths $\Gamma_{\tau \nu}^c(s^0)$, our notation will most times be simplified to 
$\gamma_{\tau \nu}^{c \, \sigma}$, 
where $c_\sigma \subset \bar{c}_\tau$ is the cell in $C$ corresponding to $s^0 \in {\cal N}_C^0$. 
An extended lattice gauge field 
takes a $k$-simplex of combinatorial paths and 
assigns to it the relative homotopy class of a $k$-simplex in $G$: 
\begin{equation}\label{Ac}
A^c(\gamma_{\tau \nu}^{c \, \sigma}) \in G , 
\quad 
A^c(\Gamma_{\tau \nu}^c(s^k)) \in {\mathrm GlExt}_k 
\quad \mbox{ for } 
1 \leq l \leq n-1 
\end{equation}
where $A^c( \Gamma_{\tau \nu}^c(s^k) ) \in {\mathrm GlExt}_k$ 
is the homotopy class of an $k$-dimensional 
hypersurface with boundary and corners in $G$ 
relative to fixed points obtained from 
the evaluation of $A^c$ 
on the vertices of 
$s^k$ and also relative to the evaluation 
$A^c( \Gamma_{\tau \nu}^c(s^m) ) \in {\mathrm GlExt}_m$ 
for all the lower dimensional subsimplices $s^m \subset \partial s^k$. 
The set of evaluations of $A^c$ must obey a set of consistency conditions: 
\begin{itemize}
\item
For any triple of 
nested closed cells 
$\bar{c}_\nu \supsetneq \bar{c}_\tau \supsetneq \bar{c}_\sigma$ 
and any $0$-simplex 
contained in the smallest closed cell 
$c_{\rho} \subset \bar{c}_\sigma$, 
there is a compatibility condition 
\begin{equation}\label{Cocyclv}
A^c( \gamma_{\sigma \nu}^{c \, \rho} ) = 
A^c( \gamma_{\sigma \tau}^{c \, \rho} ) 
A^c( \gamma_{\tau \nu}^{c \, \rho} ). 
\end{equation}
This condition is natural when regarding the gauge field as providing gluing maps forming a \v{C}ech cocycle, and it is equivalent to demanding that the evaluation of 
$A^c$ in the given set of paths induces a groupoid homomorphism from the combinatorial path groupoid to $G$. 
An alternative is to define $A^c$ as providing 
parallel transport maps on a 
set of independent generators of the combinatorial path groupoid 
$A^c ((\gamma_\nu^{x = p_\sigma})^c) \in G$ for all pairs of 
nested closed cells $\bar{c}_\nu \supsetneq \bar{c}_\sigma$. 
\item
Given any simplex $s^k \subset \bar{c}_\tau \subsetneq \bar{c}_\nu$ 
with $2 \leq k \leq n-1$, 
construct a ($k-1$)-dimensional hypersurface in $G$ up to relative homotopy using 
the points in $G$ and relative homotopy types contained in the set 
\begin{equation}\label{GlExtInt}
\{ A^c( \Gamma_{\tau \nu}^c(s^l) ) \}_{s^l \subset \partial {s^k}} . 
\end{equation}
The result must be compatible with a contractible hypersurface 
{\em after the condition on the homotopy of maintaining 
fixed position of the points in $G$ resulting from the evaluation on 
vertices of the simplices is lifted}. 
\item
The consistency condition (\ref{Cocyclv}) 
demanded above to the $0$-dimensional simplices needs to be extendible to the interior of simplices $|s|$ of any dimension (where $|s|$ denotes the geometric realization of the abstract simplex $s$). 
The condition can be stated as demanding triviality of 
\begin{equation}\label{CocylceExtToS}
A^c( (\Gamma_{\sigma \nu}^c(s))^{-1} )  
A^c( \Gamma_{\sigma \tau}^c(s) ) A^c( \Gamma_{\tau \nu}^c(s) ) , 
\end{equation}
where several things must be explained. 
By $(\Gamma_{\sigma \nu}^c(s))^{-1}$ we mean the simplex of paths composed by the inverses of the paths in $\Gamma_{\sigma \nu}^c(s)$. 
The product of relative homotopy classes of $\dim(s)$-hypersurfaces 
in $G$ written above is defined as follows: 
For each factor in the product, 
consider a representative parametrized by points in $|s|$. 
Then calculate the product by point-wise evaluation, 
and calculate the homotopy class relative to the boundary conditions resulting from the product of the boundary conditions of the factors. 
Demanding a trivial homotopy type means that the resulting homotopy class is that of a hypersurface that can be contracted to the identity in $G$. 
\end{itemize}

A gauge transformation acts on extended lattice gauge fields by conjugation. 
The possible sources and targets of combinatorial paths are points in 
${\cal N}_C$, which correspond to cells in $C$. Each assignment of group elements to the cells of $C$ determines a gauge transformation $g$ acting on a given extended lattice gauge field $A^c$ as follows: 
\begin{equation}\label{gtransf}
(g \triangleright A^c) ( \Gamma_{\tau \nu}^c(s) ) = 
g_\tau ( A^c ( \Gamma_{\tau \nu}^c(s) ) ) g_\nu^{-1} . 
\end{equation}


Results of the previous sections imply that an extended lattice gauge field characterizes a $G$-bundle over the combinatorial base 
$|{\cal N}_C|$ up to equivalence and also determines a parallel transport map on the abstract lattice ${\cal L}_C = |{\cal N}_C^{(1)}|$. 

In Section \ref{MacroObs} we defined extended lattice gauge fields as the result of evaluation of the algebra of macroscopic observables 
${\rm Obs}_{M, C}$ defined for gauge fields in the continuum. 
The relation between extended lattice gauge fields as defined in that section and the ones defined in this subsection is 
a one to one correspondence 
given by the simplicial isomorphism between the simplicial complexes 
${\cal N}_C$ and ${\mathsf N}_C$ induced by the homeomorphism 
$\phi_C: |{\cal N}_C| \to M$ which defines the triangulation 
${\mathsf N}_C$. 
The mentioned simplicial map, which is determined by a map between the vertex sets respecting the relations which determine the higher dimensional simplices of the complex, does not depend on all the details of $\phi_C$. The images of the simplices by the homeomorphism $\phi_C$ 
allow us to assign an extended lattice gauge field to each 
gauge field in the continuum. 
This is the interpretation of the algebra of macroscopic observables 
${\rm Obs}_{M, C}$.

\subsection{Comments on regularization}
\label{CommentsReg}

We saw how the set of macroscopic observables defined in Section 
\ref{MacroObs} 
evaluated on a given gauge field in the continuum $A$ 
leads to an extended lattice gauge field $A^c$. 
We can summarize this by defining an evaluation map 
and writing 
$A \xmapsto{\tiny \mbox{e-Obs}_C}{} A^c$. 
The field $A^c$ stores our partial knowledge of the gauge field at the scale determined by a given subalgebra of macroscopic observables. 
This map tells us that we can think of extended gauge fields $A^c$ as equivalence classes of gauge fields in the continuum, which from the point of view of the macroscopic observables are indistinguishable. 

Let us imagine that we have a map $A^c \overset{{\rm rep}}{\longmapsto} A$ 
choosing a representative from each class. The map would have to satisfy 
$\mbox{e-Obs}_C \circ {\rm rep} = {\rm id}$. 
There is no explicit formula for such a map, 
and for non-abelian gauge groups 
any construction of a map of this type is expected to be complicated. 
The possible existence of such a map, however, prompts some remarks that we give below. 

Any functional of the gauge field would be regularized by 
${\rm reg}^{\rm rep} \doteq {\rm rep}^\ast$. This applies in particular to ``holonomy functionals'' 
(if only real valued gauge invariant functionals are considered we may consider composing with appropriate class functions), to the Yang-Mills action and to the topological charge. 

Let us consider the regularization of holonomy functionals. We can regularize them as 
\[
{\rm reg}^{\rm rep} (H_l) [A^c] = H_l [A(A^c)]  
\]
or by using the simpler regularization 
${\rm reg}^{\rm pwf} (H_l) [A^c] \doteq 
H_{l^c(l)} [A] = H_{l^c(l)} [A^c]$. 
Notice that for any loop $l$ such that its corresponding combinatorial loop $l^c(l)= {\mathsf T}_C(l)$ is trivial the resulting regularized holonomy functional evaluates to the identity on any extended lattice gauge field $A^c$. 
We could say that the regularization consists on evaluating on gauge fields that are flat except for conical singularities that are unnoticed by large classes of loops. 
In recent years a variation of loop quantum gravity 
using a space of gauge fields whose elements are piecewise flat connections with conical singularities has gained a lot of attention; see for example \cite{PWflat}.

In the regularization based on piecewise flat connections with conical singularities 
${\rm reg}^{\rm pwf}$, 
the information concerning the gluing extension type observables 
(\ref{gll}) is neglected. 
On the other hand, a regularization ${\rm reg}^{\rm rep}$ 
needs that information 
to produce a gauge field $A$ continuously interpolating between evaluations of gluing maps on a discrete collection of points, and the non homotopic choices of such interpolations are labeled by the gluing extension observables (\ref{gll}).

We could ask if the more complicated 
${\rm reg}^{\rm rep}$ is worth the extra effort. 
The answer is that it depends on what we are studying. 
If we want to regularize holonomy functionals, we can use 
${\rm reg}^{\rm pwf}$ and get the same outcome for a large family of loops. 
On the other hand, if we are interested in any aspect in which the continuity of the gauge field is essential, like in the case of 
the relational observable 
$[\psi_{c_\nu^c \bar{c}_\nu}]$ 
defined in Section \ref{ARelObs} (which turns out to be a topological charge when $M \simeq S^4$), then 
a regularization ${\rm reg}^{\rm rep}$ would let us 
calculate the exact value of the observable. The calculation is 
$[\psi_{c_\nu^c \bar{c}_\nu}][A^c] = [\psi_{c_\nu^c \bar{c}_\nu}[A(A^c)]]$.
which, 
since $A^c$ determines the transition function up to relative homotopy, 
is independent of the chosen representative $A(A^c)$. 

Since the role of regularization is to provide simple well-defined counterparts of kinematical observables in the continuum, it is likely that a universal regularization like ${\rm reg}^{\rm rep}$ will not be directly useful. On the other hand, it 
would provide a test indicating how good a given regularization is. 
For example, consider a regularization $S_{\tt lat}$ of the continuum action $S$ such that according to 
the corresponding measure 
the weight  $\mu_{S_{\tt lat}}({\cal B})$ assigned to a given measurable set ${\cal B}$ of combinatorial gauge fields is relatively large. 
It is desirable that for combinatorial gauge fields 
$A^c \in {\cal B}$ we have 
$S_{\tt lat}[A^c] \sim {\rm reg}^{\rm rep} (S) [A^c] = S [A(A^c)]$;  otherwise, we could not say that 
the lattice theory is a quantization of the classical theory determined by the action $S$. 
The idea of this criterion is not new; 
however, without a combinatorial gauge field $A^c$ capable of determining a $G$-bundle up to equivalence over the given base, a topological sector would have to be arbitrarily chosen.

%
%
%


In a continuum limit we would like to study the issue of reemergence of the symmetries that were destroyed by the discretization. The case of the two-dimensional Ising model is a classical example where the lattice regularization destroys translation symmetry, but in the continuum that symmetry is recovered. An explicit way of seeing this reemergence in the context of ``loop quantized theories'' was proposed in 
\cite{LQasContLim}. In that reference the continuum limit of a family of lattice theories is a loop quantized theory. The mentioned reemergence of translation symmetry can only be studied because of a regularization map analogous to the one studied in this section. Without the regularization map, the two point functions are only defined on a set of points that is much smaller than $\R^2$ (even after the continuum limit is taken). 
If the continuum limit is ever at reach in quantum gravity, we would like to study the reemergence of diffeomorphism symmetry, and 
regularization maps would have to be considered.

A question that must be asked is whether or not 
by focussing attention on 
physically motivated questions and by using the measure induced by the action 
the study would lead to different conclusions. 
A particular case when this discussion arises is the following. We showed that the relational observable defined in Section \ref{ARelObs} could be studied using extended lattice gauge fields and not using ordinary lattice gauge fields, when the physical measure was ignored. 
For a discussion of this issue in relation to the work of Lüscher in lattice gauge theory, see Section \ref{Summary+Outlook}.

\section{The space of extended lattice gauge fields}
\label{M_C}
%
The space of extended lattice gauge fields is a finite dimensional manifold 
${\cal M}_{{\mathsf N}_C}$. Its relation with the space of ordinary lattice gauge fields ${\cal M}_{{\mathsf L}_C}$ is given by a projection map 
\begin{equation}
\pi_{{\mathsf L}_C {\mathsf N}_C} : 
{\cal M}_{{\mathsf N}_C} \to {\cal M}_{{\mathsf L}_C} 
\end{equation}
describing a covering map where the discrete group $R_C(G)$ defined in Section \ref{MacroObs} acts freely and transitively on the fibers. 
This structure helps us to understand the space ${\cal M}_{{\mathsf N}_C}$. The space of ordinary lattice gauge fields can be parametrized by a group manifold 
${\cal M}_{{\mathsf L}_C} \simeq G^{N_1({{\mathsf L}_C})}$ where there is a copy of $G$ for each link in the lattice 
${{\mathsf L}_C} = {\mathsf N}_C^{(1)}$. Thus, any two lattice gauge fields differ by 
an element of $G^{N_1({{\mathsf L}_C})}$.

While the space ${\cal M}_{{\mathsf L}_C}$ is connected, 
the set of connected components of ${\cal M}_{{\mathsf N}_C}$ is 
in one-to-one correspondence with the set of equivalence classes of 
$G$-bundles over $M$. 
For a more detailed description, see \cite{ELGmath}. 

The notation ${\cal M}_{{\mathsf N}_C}$ indicates that the 
auxiliary structure used to define extended lattice gauge fields is the triangulation ${\mathsf N}_C$. 
In \cite{ELGmath} we gave a definition of extended lattice gauge fields in which the participating auxiliary structure is $C$, instead of the refining triangulation ${\mathsf N}_C$, the space of fields defined in 
\cite{ELGmath} 
is denoted by ${\cal M}_C$. In the Appendix we give a map from 
${\cal M}_{{\mathsf N}_C}$ to ${\cal M}_C$.

\section{Coarse graining the gauge field}\label{CoarseGrainingSect}
%

Consider two cellular decompositions $C, C'$ of $M$. The measuring scale determined by 
the algebra of macroscopic observables 
${\rm Obs}_{M, C'}$
is finer than the one determined by ${\rm Obs}_{M, C}$ if there is an algebra homomorphism 
\[
{\rm Obs}_{M, C} \to {\rm Obs}_{M, C'} 
\]
that is an isomorphism on its image. 
In other words, 
we need to have an expression for 
every $C$ observable in terms of $C'$ observables. 

The observables that we are discussing are 
evaluations of gluing maps on a discrete set of points 
and homotopy types of gluing maps restricted to simplices ${\mathsf s} \in {\mathsf N}_C$. 
Each of these corresponds to a relative homotopy class of 
hypersurfaces in $G$ determined by the evaluation of 
a gluing map induced by a $(\dim {\mathsf s})$-simplex of paths in the continuum. 

Let us start solving the problem with the observables 
(\ref{glv}) 
corresponding to the evaluations of gluing maps on a discrete set of points. 
There is another way to understand the same observables that solves the coarse graining problem. 
We mentioned that observables (\ref{glv}) could also be seen as characterizing a groupoid homomorphism between the groupoid of paths fitting in ${\mathsf L}_C$ to the internal gauge group, 
${{\cal P}_{{\mathsf L}_C}} \to G$. 
Thus, as far as the set of observables (\ref{glv}) is concerned, 
${\rm Obs}_{M, C} \to {\rm Obs}_{M, C'}$ is induced by 
a groupoid homomorphism 
${\mathsf R}_{C'\, C} : {{\cal P}_{{\mathsf L}_C}} \to 
{{\cal P}_{{\mathsf L}_{C'}}}$ 
which is an isomorphism on its image.

In Section \ref{RegSect} we defined the groupoid homomorphism 
${\mathsf T}_C:{\cal P}_M \to {{\cal P}_{{\mathsf L}_C}}$. 
It is clear that ${\mathsf R}_{C'\, C} = T_{C'}|_{{\cal P}_{{\mathsf L}_C}}$. 
The requirement that the groupoid homomorphism be an isomorphism into its image is met if $C' \geq C$. 

Notice that we are not using all the details of the auxiliary structure contained in the refining triangulation ${\mathsf N}_C$. The only thing that is used is a choice of groupoid homomorphism 
${\mathsf R}_{C'\, C} = T_{C'}|_{{\cal P}_{{\mathsf L}_C}}: {{\cal P}_{{\mathsf L}_C}} \to {{\cal P}_{{\mathsf L}_{C'}}}$, 
among finitely many choices (if $C'$ has finitely many cells). 
However, as the cellular decomposition $C'$ is refined 
every detail of the auxiliary structure becomes relevant.

There is a dual way of thinking about coarse graining. 
The problem is, given a pair of cellular decompositions 
$C' \geq C$, 
finding a map ${\cal M}_{C'} \to {\cal M}_C$ 
assigning an extended lattice gauge field with respect to $C$ to any extended lattice gauge field with respect to $C'$. 
In principle, one way to solve this problem is to give the following sequence of maps 
$A^c_{C'} \overset{\tiny{\rm rep}_{C'}}{\longmapsto} A 
\overset{\tiny \mbox{e-Obs}_C}{\longmapsto} A^c_C$. 
The map $\mbox{e-Obs}_C$ was given in Section \ref{MacroObs}, 
but we do not have an explicit assignment 
${\rm rep}_{C'}$ 
of gauge fields in the continuum to extended lattice gauge fields.

Coarse graining at the level of standard lattice gauge fields 
${\cal M}_{L(C')} \to {\cal M}_{{{\mathsf L}_C}}$ 
is given by the pull-back of the groupoid homomorphism 
$R_{C'\, C} : {\cal P}_C \to {\cal P}_{C'}$, which follows from 
the natural isomorphisms between the path groupoids of the abstract and embedded lattices and the groupoid homomorphism 
${\mathsf R}_{C'\, C}$ defined above. 
This coarse graining mechanism was used in the context of quantum gravity earlier in \cite{Cflat}.

Extending the coarse graining map to the gluing extension data is not as simple because 
we lack an algebraic structure analogous to the groupoid structure of paths in the case of the higher dimensional objects on which extended lattice gauge fields act. 
We can compose relative homotopy classes of 
gluing extensions, but we do not have a sufficient understanding of the rules satisfied by those compositions. 
For specific situations, like extracting gluing extension data for 
a complete cell 
$[ g_{\tau \nu} [A]]$ 
from the recorded data corresponding to its restrictions to subsimplices, coarse graining is trivial. Another specific situation where coarse graining is trivial, but important for our framework, is in the statement of compatibility condition (\ref{GlExtInt}) for $A^c$. That condition requires the calculation of the relative homotopy class of a hypersurface in $G$ constructed by gluing pieces associated with simplices in the boundary of a given simplex. Those situations involve a type of ``local coarse graining'', where gluing the set of pieces does not involve any choices. 
We have successfully studied a few examples of 
coarse graining involving choosing different orders, or even the use of different types of gluing, but we have not yet finished understanding the general picture.

Coarse graining is a crucial process when working with effective field theories in the context of Wilsonian renormalization. It is also relevant if we are interested in studying the coarsest aspects of the gauge field: the bundle structure induced by the gauge field. Consider, for example, the case of $SU(2)$ gauge fields over $S^4$. We have described how an extended lattice gauge field determines the bundle because it characterizes the transition functions between overlapping local trivializations. 
If the cellular decomposition $C$ has only two $4$-cells the bundle structure is easily read from the transition function $\psi_{\nu \, \mu}$ defined on the intersection of the two charts 
$\bar{c}_\nu \cap \bar{c}_\mu \simeq S^3$. In particular, the topological charge is the winding number of the induced map 
$\psi_{\nu \, \mu}: S^3 \to SU(2)$.

If one of the $4$-cells in $C$ is the coarse graining of a finer cellular decomposition $C'$, then the situation described above arises from the coarse graining of the finer cellular decomposition. The coarse graining of parallel transport necessary for providing gluing data on vertices follows the pullback of the groupoid homomorphism 
${\mathsf R}_{C'\, C}$. 
Since we are considering $G = SU(2)$ (and not $G= SO(3)$), 
the extension data for $1$ and $2$-dimensional simplices are calculated directly from the parallel transport data without the need for any topological input. Only the gluing extension data for dimension $3$ simplices is nontrivial. It is the relative homotopy class of a $3$-dimensional hypersurface in $SU(2)$, which is calculated from the relative homotopy class of smaller pieces of the surface. The relative homotopy type of such smaller pieces of the surface can be extracted from the extended lattice gauge field $A^c_{C'}$, and then it can be glued to obtain gluing extension data at scale $C$. 
The extraction of gluing extension data at scale $C$ from $A^c_{C'}$ 
is not a straightforward addition of integers, and coarse graining can be done following many different gluing orders. The set of 
relative homotopy classes 
${\mathrm GlExt}_3$ with its gluing rules in the case $G=SU(2)$, however,  
behaves in an abelian way.

In \cite{ELGmath} we study another way to relate gauge fields defined on different cellular decompositions (which are dual to triangulations). 
Any two cellular decompositions of $M$ that are both dual to triangulations are related by a finite sequence of dual Pachner moves. 
Two cellular decompositions $C, C'$ are related by a dual Pachner move 
if there is a closed $n$-disc $D \subset M$ 
such that $C|_{M \setminus D} = C'|_{M \setminus D}$ and that 
inherits a cellular decomposition from $C$, called $H^S$, 
and another one from $C'$, called $H^N$, which agree on $\partial D$. 
It is clear that 
$H^N$ can be glued to a copy of $H^S$ with the orientation reversed to form a cellular decomposition of $S^n$. 
The Pachner move corresponds to deleting the cellular decomposition of $D$ associated with the southern hemisphere $H^S$ and replacing it with the cellular decomposition carried by $H^N$. 
We can describe a transition from a $A^c_C$ to $A^c_{C'}$ 
in terms of an extended lattice gauge field 
$A^c = A^c|_{H^N} \#_{\partial D} A^c_{\bar{H}^S}$ on $S^n$. 
In \cite{ELGmath} we see that 
the bundles induced by 
$A^c_C$ and $A^c_{C'}$ are equivalent if and only if 
the extended lattice gauge field 
$A^c = A^c|_{H^N} \#_{\partial D} A^c_{\bar{H}^S}$ 
induces a trivial bundle over $S^n$. 

Pachner moves have the disadvantage of not being generically coarse graining nor refining. 
Recently Dittrich and Geiller introduced 
a formulation of canonical quantum gravity where the excitations of the field are modeled by conical singularities of the gauge field, and therefore they can be located at codimension two simplices of a triangulation. Star subdivision moves (or Alexander moves), which are always refining, were used as a primary ingredient in taking their framework to the continuum 
\cite{PWflat}. 
In our case we could study dual star subdivision moves if we restrict to cellular decompositions that are dual to triangulations. The resulting coarse graining scenario, however, does not fit naturally with what we described above.

Another interesting approach to coarse graining in canonical loop quantum gravity was proposed by Livine; see \cite{EteraCoarse} and references therein. 
It shares some elements with \cite{PWflat} that are not shared with our proposal. As mentioned previously, that two scales are related by one being finer than the other is ultimately a statement about algebras of macroscopic observables. The two proposals mentioned above do not use holonomies as the basis of their algebra of macroscopic observables as we do: they use fluxes. 
Despite this difference, the two mentioned coarse graining proposals and ours share some elements in their construction. 
In Section \ref{RegSect} we described a groupoid homomorphism between the path groupoid in the continuum and the groupoid of combinatorial paths. The assignment was defined in terms of the sequence of cells induced by a path in the continuum; however, it is not hard to 
phrase the definition of the groupoid homomorphism 
in terms of the intersection type of the path in the continuum with the array of closed codimension one cells. 
In the context of loop quantum gravity, one may think of an array of surfaces (which could be given by the codimension one cells of a cellular decomposition) 
in space measuring the flux across those surfaces. These are kinematical observables in loop quantum gravity related to the measurement of area and related to the spin network basis. This set of observables play the primary role in 
\cite{EteraCoarse} and in \cite{PWflat}. 
A careful study of the relation between their proposal and ours may be interesting. The subtle part is to have good definitions of subalgebras of observables in loop quantum gravity. If the mentioned collection of surfaces is composed by open $2$-cells that do not intersect, the induced collection of area operators would commute, but in the places where there are $1$-cells of the cellular decomposition it would have ``holes.'' On the other hand, choosing an array of closed $2$-cells which intersect at links of the cellular decomposition, has the disadvantage of leading to area operators that do not commute. 
Another comment regarding the relation between our work and the articles mentioned above is that in 
all of them some choices are necessary in defining the coarse graining maps, and the choices are of 
the same type as the ones involved in our construction. They resemble the choices described above to define the map 
$R_{C'\, C} : {\cal P}_C \to {\cal P}_{C'}$.

\section{Summary and outlook}
\label{Summary+Outlook}
%
%
%

The concept of an extended lattice gauge field was recently introduced in 
\cite{ELGmath}. Here we see how this concept arises naturally from the set of evaluations of the ``ensemble of local subalgebras of macroscopic observables'' defined in Section \ref{MacroObs}. In this scenario gauge fields in the continuum break down into different equivalence classes labeled by the evaluation of a set of macroscopic observables determined by a cellular decomposition $C$ of $M$ together with some extra auxiliary structure. 

These observables arise from the decimation of a collection of maps gluing local trivializations of the bundle induced by the gauge field. 
The evaluation of the mentioned gluing maps on a discrete set of points is complemented by recording 
the homotopy classes of their restriction to a collection of simplices of a refining triangulation 
(relative to certain fixed structures in their boundary). 
Once we have complemented the set of observables, their evaluation 
characterizes a $G$-bundle over $M$ and its restrictions to any submanifold inheriting a cellular decomposition from $C$. 
In addition, the evaluation also 
determines the parallel transport along paths that fit in the embedded lattice ${{\mathsf L}_C}$.  

The space of extended lattice gauge fields is a finite dimensional smooth manifold covering the space of standard lattice gauge fields 
$\pi : {\cal M}_{{\mathsf N}_C} \to {\cal M}_{{\mathsf L}_C}$, 
and its 
set of connected components parametrizes 
the set of equivalence classes of $G$-bundles over $M$. 

In Section \ref{RegSect} we gave an alternative definition of extended lattice gauge fields over a base constructed as an abstract 
simplicial complex ${\cal N}_C$, and we studied the regularization of kinematical observables from a continuum gauge theory to the corresponding gauge theory over a combinatorial base. 

We also studied coarse graining of the extended lattice gauge field in Section \ref{CoarseGrainingSect}. The coarse graining of the lattice gauge fields is given by the pullback of a groupoid homomorphism 
${R}_{C'\, C} : {\cal P}_C \to {\cal P}_{C'}$ 
determined by 
intersection properties of an embedding of ${{\mathsf L}_C}$ with $C'\geq C$. 
Coarse graining of the data assigned by the extended lattice gauge field to higher dimensional objects is understood at the level of exploring examples, and we are working to gain a deeper understanding.

Let us discuss the possible physical relevance of extended lattice gauge fields. First we briefly comment on the case of lattice gauge theory for chromodynamics 
to provide elements for the discussion in the case of 
quantum gravity. 
In quantum chromodynamics the topological susceptibility is known to play a role in the calculation of the masses of hadrons. However, we mentioned that lattice gauge fields were incapable of storing topological information leading to a puzzle. 
Lüscher gave an answer to this puzzle \cite{Luscher}. He proved that 
if we are only concerned with the continuum limit, and under the assumption that the limiting process requires taking the bare coupling constant to zero, 
some topological properties of the gauge field are recovered in the limit. 
For example, the space of lattice gauge fields ``becomes disconnected in the continuum limit'' with the strata labeled by a topological charge. 
The quotation marks mean that for any lattice with any value of the coupling constant 
the space of lattice gauge fields continues to be connected, and that the actual statement is subtle. 
Lüscher's construction removes a set of fields 
from the space of allowed fields, and then he 
assigns a bundle to every allowed lattice gauge field. 
The set of removed gauge fields may be of measure zero, but it is arbitrary: 
one may imagine removing a different set of measure zero (which would look somehow unnatural) 
and defining a corresponding assignment of allowed fields to bundles. 
In that situation there would be a set of fields for which the two procedures yield inequivalent bundles. 
The measure of the mentioned set of fields, however, decreases and asymptotically vanishes 
as the coupling constant becomes smaller. 

If one is interested in topological features of the gauge field at a lattice spacing different from zero, our extension does provide new possibilities because in that scenario 
the macroscopic observables (\ref{gll}) become physically relevant. 
For example, one may explore the possibility of using extended lattice fields to describe effective theories at macroscopic scales and explore topological issues as the gauge field is more and more coarse grained (see Section 
\ref{CoarseGrainingSect}) 
eliminating details of the gauge field 
that are irrelevant for a given topological issue of interest.

Now let us address the possible relevance of extended lattice gauge fields in quantum gravity. 
In relation to Lüscher's work cited above, notice that 
in the case of quantum gravity the continuum limit is not expected to be determined by a gaussian fixed point of the coupling constant \cite{AsymptSafety}. Thus, results of ordinary lattice gauge theory may have to be revised before they are imported to quantum gravity. 
In addition, notice that the use of lattice gauge fields in quantum gravity, at the exploratory level at which it is currently performed, 
the physical interpretation many times does not take place in the continuum limit. For example, one argument justifying that spin foam models are indeed a quantization of general relativity 
is based on the asymptotic behavior of the amplitude for one single $4$-simplex. 
As mentioned above, if the gauge field over a discretized base is used to describe an effective theory, the macroscopic observables (\ref{gll}) become physically relevant. 
At a given scale, away from a possible continuum limit in which the bare coupling constant goes to zero, an observable like the one exhibited in Section \ref{ARelObs} needs the macroscopic observables (\ref{gll}).

A spin foam study of euclidian gravity and BF theory in two dimensions 
by Oriti, Rovelli and Speziale 
revealed that an extension of the lattice gauge field was essential in capturing the correct physics \cite{2dSF}. 
Our definition of extended lattice gauge fields could be seen as a higher dimensional non-abelian generalization of the extension used in \cite{2dSF}. An application of our extended lattice gauge fields to two-dimensional spin foam models trivially reproduces part of their results. 
It is more relevant to mention that the same topological mechanism making the extension of the gauge field relevant for euclidian two-dimensional gravity makes it potentially relevant for four-dimensional gravity. What we can prove is that 
if we would like to give the interpretation to the histories used in a discrete approach to quantum gravity 
as describing fields in the continuum up to ``microscopical details'' where the details may be regarded as homotopies of the field, then our extension of extended lattice gauge fields includes 
data that is not included in the standard approach to gauge fields on a lattice. This applies to any approach based on gauge groups for which the third homotopy group is not trivial 
($\pi_3(SO^+(3,1)) = \pi_3(SO(3)) = \pi_3(SU(2)) = \pi_3(SL(2,\C)) = \Z$). 
Another recent development in the context of three-dimensional gravity in bounded domains shows that fields with nontrivial winding numbers play a crucial role \cite{Winding3dB, Winding3dW}. 

In the case of two-dimensional gravity, one may argue that 
there is no dynamics (when each history is correctly modeled), and that this explains the relevance of topological features. 
Similarly, 
in three-dimensional gravity the dynamics freezes all local degrees of freedom, leaving a space of physical fields determined by global features; this may be the explanation of the relevance of topological features. 
In contrast, it may happen that the same topological features are completely hidden by the rich local dynamics of four-dimensional gravity. 
At this moment we do not have an argument against this scenario, but the alternative is a picture in which at different scales we see more and more details dressing a coarse topological history containing no more than the most basic topological features of the field. If a version of this scenario is realized, our framework would be relevant for four-dimensional gravity. 
At the moment there is no model for four-dimensional quantum gravity using the gluing extension data defined in (\ref{gll}).

\section*{Appendix}

In \cite{ELGmath} we presented extended lattice gauge fields in a more mathematically rigorous manner. In this article we include material with physical motivation which does not appear in that reference, and this new material can be expressed more naturally using a refining triangulation ${\mathsf N}_C$ of the cellular decomposition $C$ which does not appear in \cite{ELGmath}. The resulting extended lattice gauge fields used in this article correspond to a finer 
measuring scale than those defined in the mentioned reference. Below we give a map 
\begin{equation}
\pi_{C \,{\mathsf N}_C} : {\cal M}_{{\mathsf N}_C} \to {\cal M}_{C} 
\end{equation}
from the space of extended lattice gauge fields defined in terms of 
the refining triangulation ${\mathsf N}_C$ to the 
space of extended lattice gauge fields defined in terms of $C$. 

The first difference is that the embedded lattice 
${{\mathsf L}_C}= {\mathsf N}_C^{(1)}$
used in this article 
is finer than the ``cellular network'' $\Gamma(C)$ 
used in \cite{ELGmath}. 
Recall that ${{\mathsf L}_C} = {\mathsf N}_C^{(1)}$ 
is a collection of paths 
generated by paths 
of the type $\gamma_\nu^{x=p_\sigma}$ 
for all the pairs cells $c_\nu , c_\sigma \in C$ such that 
$\bar{c}_\nu \supsetneq \bar{c}_\sigma$. 
If we erase 
from the set of generators 
the paths $\gamma_\nu^{x=p_\sigma}$ 
in which the dimension of $c_\sigma$ is bigger than cero, 
we obtain $\Gamma(C)$. 
Fields in ${\cal M}_{C}$ only store parallel transport along paths of 
$\Gamma(C)$ and not the parallel transport along the rest of the paths in ${{\mathsf L}_C}$. 
The second difference is that in this article each closed cell 
$\bar{c}_\tau \in C$ has a refining triangulation composed by simplices 
${\mathsf s} \in {\mathsf N}_C$ with ${\mathsf s} \subset \bar{c}_\tau$. 
In Theorem 4 of \cite{ELGmath} an extended lattice gauge field is characterized in terms of the relative homotopy classes of gluing maps. 
The information recorded in an extended lattice gauge field 
in ${\cal M}_{C}$ is the 
homotopy class of each gluing map $[g_{\tau \nu}|_{\bar{c_\sigma}}]$ 
(for all triples of nested closed cells 
$\bar{c}_\nu \supsetneq \bar{c}_\tau \supset c_\sigma$) 
relative to its evaluation on the vertices of $\bar{c}_\tau$, 
which is determined by the parallel transport along paths of 
$\Gamma(C)$, and also restricted to be compatible with 
similar homotopy data related to all the lower dimensional cells 
$c_{\sigma'} \subset \partial c_\sigma$. 

Given $A^c \in {\cal M}_{{\mathsf N}_C}$ we give below 
the corresponding relative homotopy classes 
$[g_{\tau \nu}|_{\bar{c_\sigma}}]$ 
characterizing an extended lattice gauge field in ${\cal M}_{C}$. 
\begin{itemize}
\item
If $\dim(c_\sigma) = 0$ 
the meaning of $[g_{\tau \nu}|_{\bar{c_\sigma}}]$ 
is simply evaluation of the gluing map at the zero-dimensional cell, i.e. vertex, $c_\sigma \in C$. This evaluation is 
determined by $A^c$ as defined in (\ref{glv}). We can write 
\[
[g_{\tau \nu}|_{\bar{c_\sigma}}] = 
[g_{\tau \nu}|_{\bar{c_\sigma}}] [A^c]. 
\] 
\item
If $\dim(c_\sigma) = m > 0$, then (\ref{gll}) says that $A^c$ determines the collection of relative homotopy classes 
$[g_{\tau \nu}|_{{\mathsf s}^m}] [A^c]$ 
for all $m$-simplices ${\mathsf s}^m \in {\mathsf N}_C^m$. 
The relative homotopy class $[g_{\tau \nu}|_{\bar{c_\sigma}}]$ 
induced by an extended lattice gauge field in ${\cal M}_{C}$ 
is obtained by considering any 
continuous $m$-dimensional hypersurface which gives a 
representative of 
each class in the collection 
$\{[g_{\tau \nu}|_{{\mathsf s}^m}] 
[A^c]\}_{{\mathsf s}^m \subset \bar{c_\sigma}}$ 
and calculating its homotopy class relative to the smaller set of restrictions imposed by $C$. We may write again 
\[
[g_{\tau \nu}|_{\bar{c_\sigma}}] = 
[g_{\tau \nu}|_{\bar{c_\sigma}}] [A^c]. 
\] 
\end{itemize}

\section*{Acknowledgements}
Claudio Meneses was supported by the DFG SPP 2026 priority programme ``Geometry at infinity". 
José A. Zapata was supported in part by grant 
PAPIIT-UNAM IN100218.


\medskip
 
\bibliographystyle{unsrt}
\bibliography{ELGFqg}{}

\end{document}